\documentclass{article}
\usepackage[utf8]{inputenc}
\usepackage{graphicx}
\usepackage[margin=2.5cm]{geometry}
\usepackage[affil-it]{authblk}
\usepackage{subfigure}
\usepackage{url}
\usepackage{amsmath}

\graphicspath{{images/}}

\title{A distributed memory, local configuration technique for re-configurable logic designs}
\author[1]{Alexander E. Beasley\footnote{corresponding author: Alexander Beasley, alex.beasley@uwe.ac.uk} }
\affil[1]{Unconventional Computing Laboratory, UWE, Bristol, UK}

\date{\today}

\begin{document}

\maketitle

\begin{abstract}
    \noindent
    The use and location of memory in integrated circuits plays a key factor in their performance. Memory requires large physical area, access times limit overall system performance and connectivity can result in large fan-out. Modern FPGA systems and ASICs contain an area of memory used to set the operation of the device from a series of commands set by a host. Implementing these settings registers requires a level of care otherwise the resulting implementation can result in a number of large fan-out nets that consume valuable resources complicating the placement of timing critical pathways. This paper presents an architecture for implementing and programming these settings registers in a distributed method across an FPGA and how the presented architecture works in both clock-domain crossing and dynamic partial re-configuration applications. The design is compared to that of a `global' settings register architecture. We implement the architectures using Intel FPGAs Quartus Prime software targeting an Intel FPGA Cyclone V. It is shown that the distributed memory architecture has a smaller resource cost (as small as 25\% of the ALMs and 20\% of the registers) compared to the global memory architectures.   

\end{abstract}

\section{Introduction}
The use of memory, memory accessing, and memory mapping techniques has a large impact on system performance~\cite{Tirri1995hardaddress,Cordasco2007boundedcollision,Rutgers2013manycores}. Efficient mapping techniques, reduction in communication overhead and the use of distributed memories can vastly increase the systems overall performance, particularly for intensive tasks such as loops and scalable graph operations~\cite{Darte1993uniformloops,Kontothanassis1996memorymap,Lin2013fastscalablegraph,Azad2016bipartite,Lo1991graph}.
The implementation of memories inside an embedded system comes with many research possibilities. Memory technologies are becoming denser and faster, allowing for higher density memory to be implemented close to its point of use. Despite this, memory still requires large, physical space. \par

Distributed memory, where the memory is close to the point at which it is used offers huge benefits, so long as the memories are kept coherent where necessary~\cite{Choi1997distributed,Klein2014billions,Huang1996os}. \par

Integrated circuits often require memory to store user defined settings that control the mode of operation. Such examples could be the sample rate or resolution of a ADC; the applied phase shift of an RF phase shifter; the gain of a variable gain amplifier and so on.\par

Field Programmable Gate Arrays (FPGAs) provide a flexible platform for designers to fabricate seemingly endless weird and wonderful systems. Quite often designers wish to make parameterisable systems where their operation can be controlled by based on a number of settings. One way to achieve this is by use of parameters~\cite{VerilogParameters} (or generics --- VHDL~\cite{VhdlGenerics}) that can be set at compile time. Parameters (generics) are a very powerful tool available in hardware descriptive languages, to create re-useable code. However, each combination of settings must be compiled separately, introducing a large amount of processing overhead and leading to a separate image file per configuration. Alternatively, designers can implement an area of the FPGA as an array of registers in which settings can be stored and propagated across the design. These registers can be programmed by means of a connection with a host system (typically USB in a modern system). Implementing these features in a device create demand on resources, reducing the overall resources available to be used as functional logic.\par

FPGA place and route stages are complicated procedures, attempting to locate resources as close as possible to reduce routing complication and net delay~\cite{Fobel2009Hardwareaccelerated,Wrighton2003simulatedannealingplacement,Gudise2004PSOplacement,Haldar2000ParallelPlacement}. As the resources are fixed, this often results in trade-offs between quality of the fitter result and run-time of the fitter~\cite{Mulpuri2001runtimetradoffs}. In addition, large fan out nets often take priority during the `fitting' stage of an FPGAs compilation. Synthesis tools often attempt to insert extra resources or promote high fan-out nets to the clocking nets~\cite{AN903TimingClosure,XilinxTimingClosure}; reducing the available resources for timing critical pathways. This leads to more complicated designs that suffer from bottle-necking, manifesting itself as a reduction in the maximum operating frequency of a design.\par

FPGAs have a large amount of memory distributed throughout the device. This memory neatly lends itself for tasks where distributed memory, close to the point of use, such as loop operations and array intensive operations~\cite{Pal2007distributedmemorysynthesis}. By extension, we can use the embedded memory blocks to create the sets of registers used to set up and control the FPGA. Distributing the settings across the FPGA to their point of use helps to reduce the required routing resources, limit the high fan-out nets and improve timing closure.\par 

In this paper we explore how a typical `global' register map expands with the number of required settings for a design and the width of each of these settings. The global register map is connected to modules in which a varying number of entries in the global register map are used to help us model the resource requirements when fanning out the register map. A second architecture is presented that removes the global register map and distributes the settings across the design to the points at which they are used. Discussions are had as to how these architectures deal with the common problem of clock domain crossing and a more recent problem of how to deal with dynamic partial reconfiguration --- the process by which a small portion of a design is changed at run-time without effecting the operation of the rest of the device.\par

The rest of this paper is organised as per the following. Section~\ref{sec:architecture} presents architectures for creating and distributing settings using a `global' register map and an architecture for distributing the settings across the device in a method that is robust to multiple clocking domains and partial dynamic re-configuration. Metrics for the architectures are presented in Sect.~\ref{sec:results}. Finally, conclusions are drawn in Sect.~\ref{sec:conclusions}.\par

\section{Architecture}
\label{sec:architecture}

\begin{figure}[!tpb]
    \centering
    \includegraphics[width=0.5\textwidth, trim=4 4 4 4,clip]{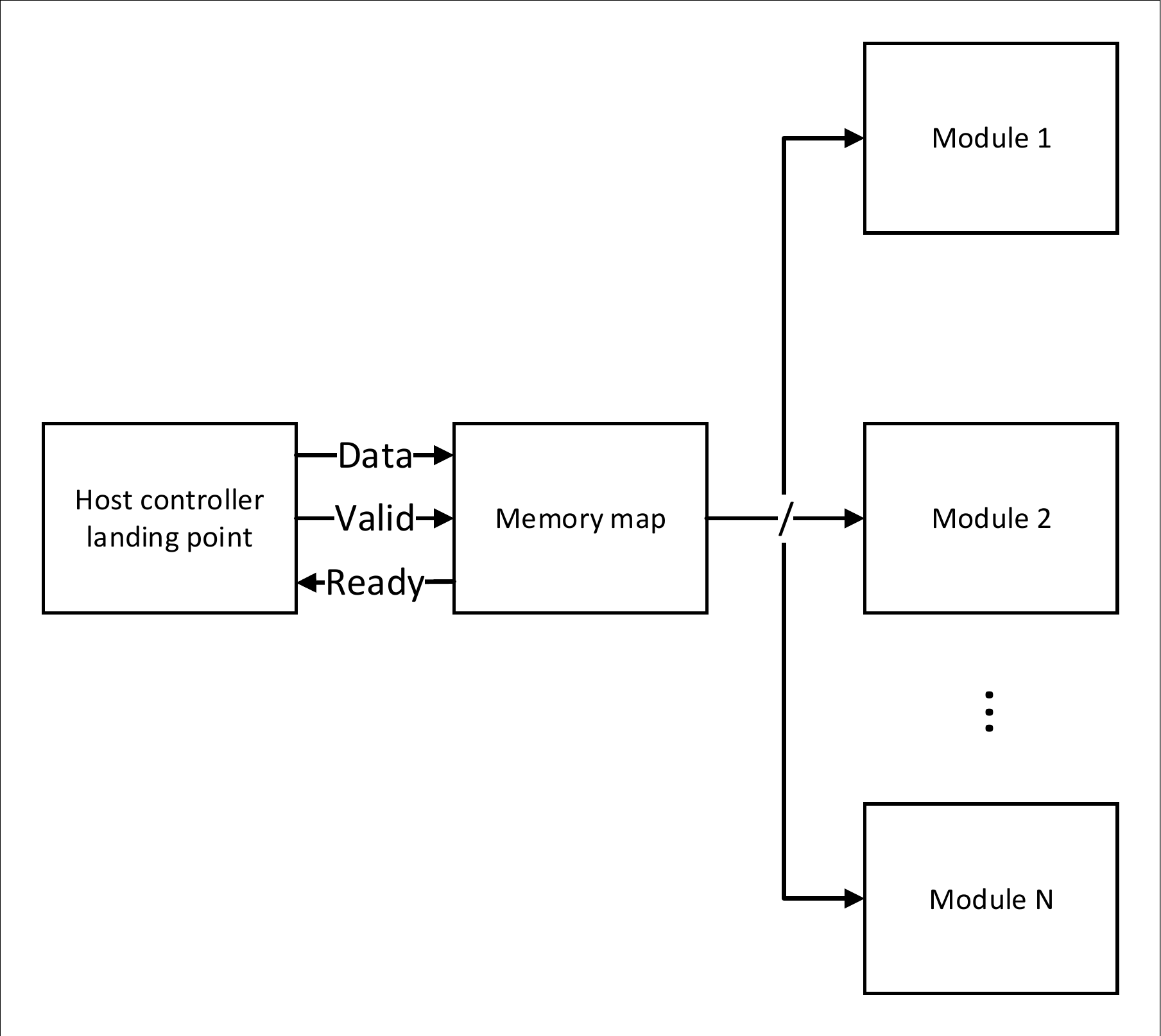}
    \caption{A global copy of the memory map is populated via the host controller, respective settings are routed to the appropriate modules on multi-bit busses.}
    \label{fig:global_arch}
\end{figure}

The settings registers are usually considered as an area of memory, in which the stored values represent modes of operation for a design. These stored values are used throughout the design to influence operation. There are a number of ways to achieve the desired behaviour, the seemingly obvious is to simply reference the values, stored in a global memory location, throughout the respective parts of the design, leading to a routing as in Fig.~\ref{fig:global_arch}.\par

\begin{figure}[!tpb]
    \centering
    \includegraphics[width=0.5\textwidth, trim=4 4 4 4,clip]{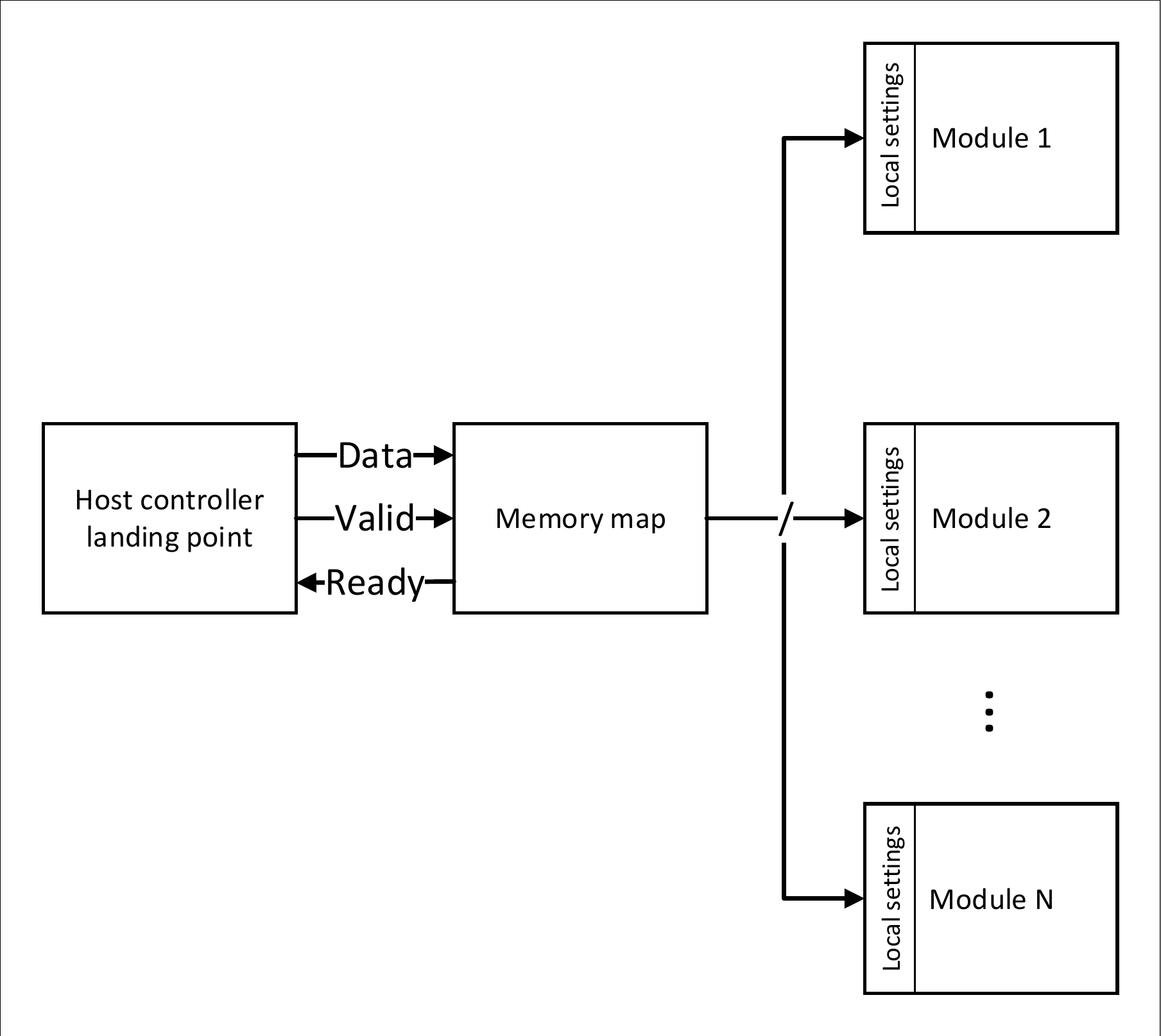}
    \caption{Entries from the global memory map are copied to where they are required locally.}
    \label{fig:global_arch_local_copies}
\end{figure}

Alternatively, distributing the memory map throughout the design moves the settings closer to where they are used. The result of which is to reduce the complexity of the routing, but not necessarily reduce the overall resource requirement. Local copies of the register map close to the point of which they are used allows the designer to safely register the values into the appropriate clock domains. The additional flip-flop stages play an important part in breaking up the total routed path into smaller elements, the shorter the path, the easier it is for a design to meet timing closure. However, the additional flip-flops used increase the overall resource cost for a design. An example of such a design can be seen in Fig.~\ref{fig:global_arch_local_copies}.\par 

\begin{figure}[!tpb]
    \centering
    \includegraphics[width=\textwidth, trim=4 4 4 4,clip]{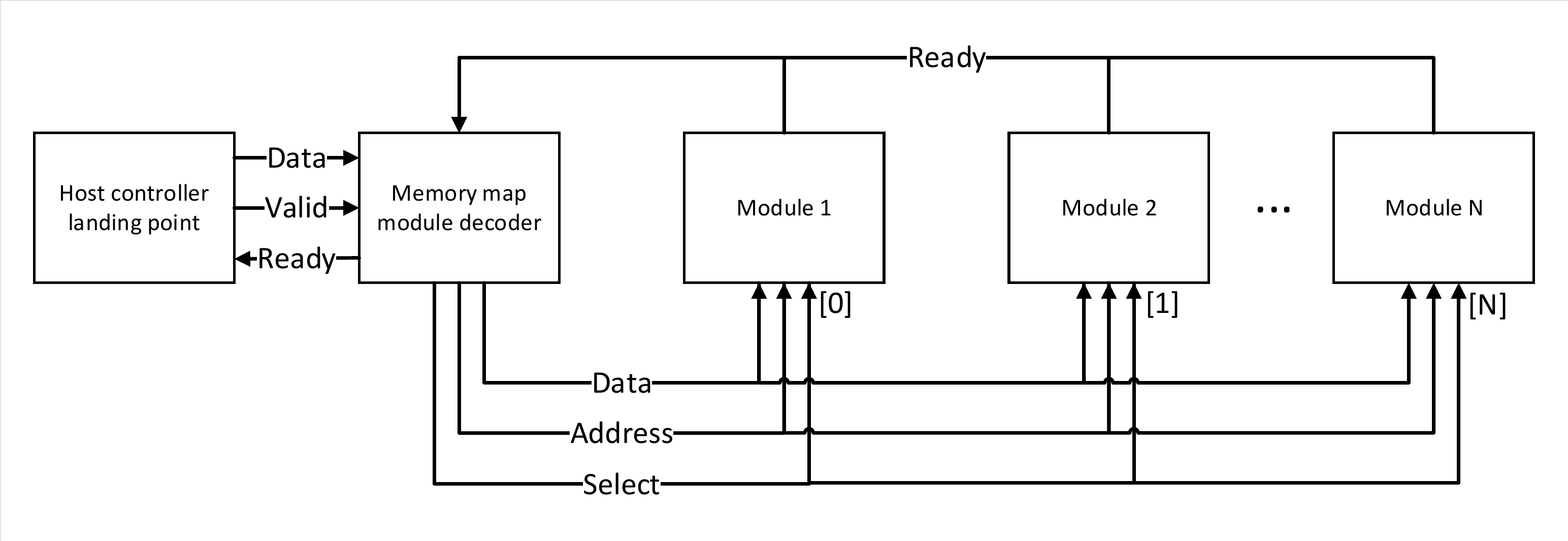}
    \caption{Bus connects elements of the design to a decoding module that distributes memory map information across the device. Uniform bus allows connection of partially dynamically re-configurable modules into the memory map bus.}
    \label{fig:distributed_arch_1}
\end{figure}

Distributing the memory map across the design can be achieved without the need for a increasing the routing complexity. Designing the distributed memory map with a common bus interface for its configuration, Fig.~\ref{fig:distributed_arch_1}, reduces the overall resource cost and significantly reduces the required routing resource.\par

The common bus interface has a number of benefits: reduced routing complexity, safe crossing into different clock domains, reduction in global memory resources, connection into dynamically partially re-configurable logic space. \par

\subsection{Clock domain crossing}

It is not uncommon for a modern digital system to use multiple clocks~\cite{Ragheb2018multipleclockdomains}, in which data are moved from one clock domain to another and memories are connected to different clock domains. Moving from one clock domain to another requires the use of safe clock domain crossing domains - which in themselves are a large research field~\cite{Li2010synchtech,Bartik2018clockdomaincrossing, Matsuda2011cdc,Preetam2015CDC} - however they require using up yet more valuable resources. Typically configuration data would be set in a slow clock domain and moved into much faster domains - potentially as very wide, parallel busses.\par

In addition to the increase in resources required for crossing clock domains, multi-clock systems lack determinism which causes problems for the verification process. Rectifying the non-deterministic nature of such systems and providing verification techniques (both stand-alone and built-in) is a rich source of research~\cite{Su2010multiclock,Leong2010cdc}. Additionally, frameworks for performing timing analyses and signal integrity in a CDC application~\cite{Matsuda2011cdc,Preetam2015CDC} have been proposed. \par

The architecture presented here, fig.~\ref{fig:distributed_arch_1}, exports a `Ready' signal from each of the subsystems. The `Ready' signal is used to indicate that the logic has been moved to a safe state in which the local memory map may be written to using the configuration bus. No changes are made to the local configuration memories while logic is operating, hence there is no danger of the registers being sampled while they are transitioning and the clock domains are safely crossed.\par

\subsection{Dynamic partial reconfiguration}

Dynamic reconfiguration and Dynamic Partial Reconfiguration (DPR) is rapidly growing in popularity as it enables FPGA designs to be changed at run-time to better meet changing systems demands~\cite{Lie2009dpr,Di2012DPRflow}. The use of DPR is rapidly gaining popularity over a number of sectors including: fault recovery~\cite{Alkady2015dprfaultrecovery}, memory controllers~\cite{Salah2017dprmemorycontroller}, real-time signal processing~\cite{Feilen2011dprrealtimesigprocessing}, software defined radio~\cite{Sadek2015dprsdr,Hosny2018dprsdr,Hassan2015dprsdr}, cognitive radio~\cite{Lie2009dprcognitiveradio}, bandwidth reduction~\cite{Najmabadi2016dprbandwidthreduction}, video filters~\cite{Khraisha2010dprvideofilter}, and RADAR signal processing~\cite{Zhang2016dprradarprocessing} to name a few.\par

\begin{figure}[!tpb]
    \centering
    \includegraphics[width=0.5\textwidth, trim=4 4 4 4,clip]{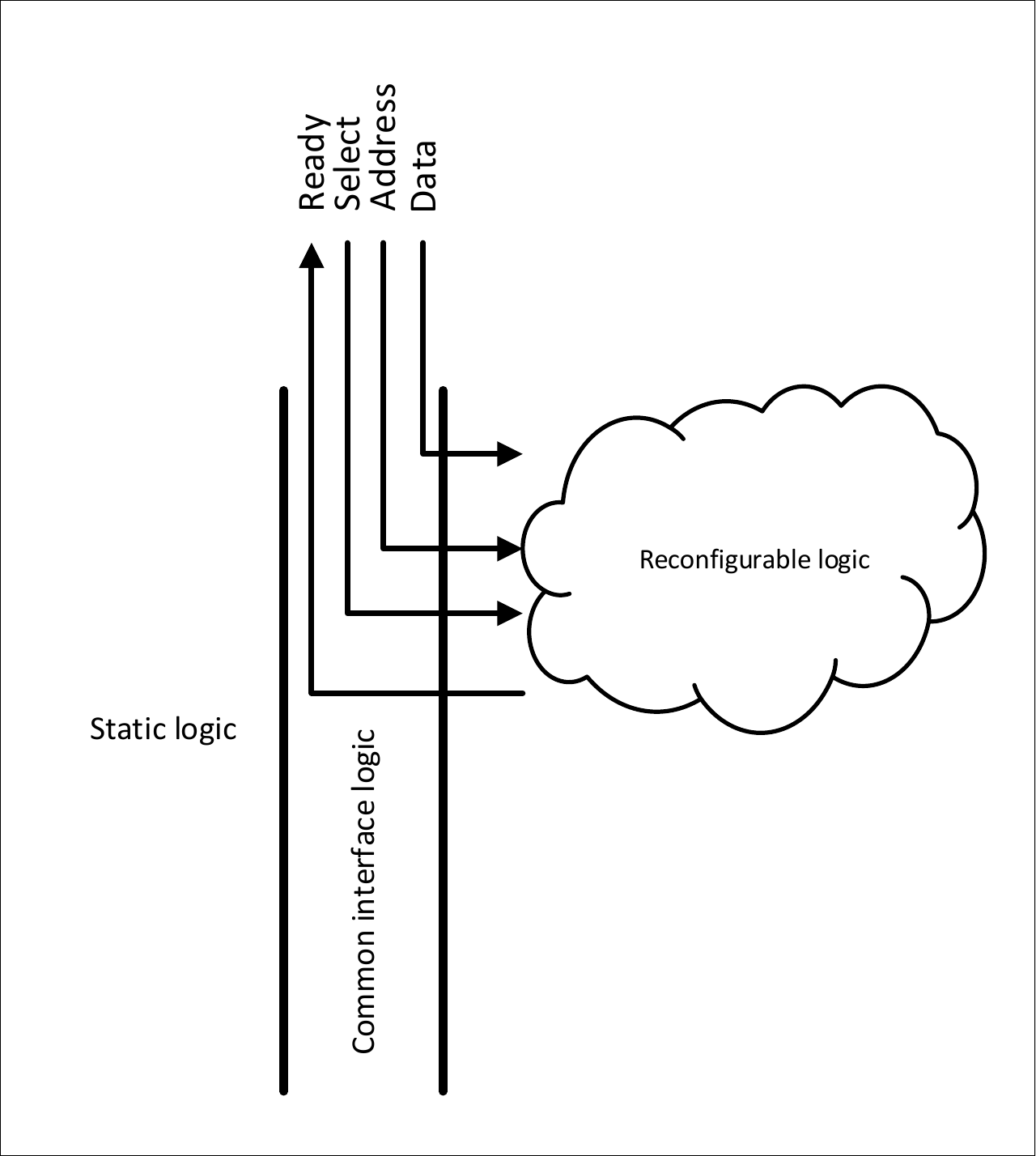}
    \caption{Partially re-configurable design showing the common programming interface in the interconnect logic between static and re-configurable logic}
    \label{fig:reconfig}
\end{figure}

DPR designs contain a mix of static logic and re-configurable logic. Between the elements of the design a common interconnect is implemented, Fig.~\ref{fig:reconfig}. The interconnect fabric contains the signals required for the configuration bus. When a module(s) in a re-configurable portion of the FPGA is changed, the configuration bus is connected into the new module along with all other data-path signals. Any settings registers inside partially configured module are then set over the configuration bus.\par

\section{Results}
\label{sec:results}

Example designs of the above architectures were written using SystemVerilog (IEEE 1800) and processed using Intel FPGA Quartus Prime 19.1.0 (Build 670); target device for compilation is a Cyclone V (5CSXFC6D6F31C8). Synthesis metrics --- Adaptive Logic Modules (ALMs), registers, combinatorial Adaptive Look Up Tables (ALUTs) and maximum operating frequency --- are presented for each architecture. Implementations are given for a variety of memory depths and widths.\par

\subsection{Global configuration - no targets}
\label{sec:global}
\begin{figure}[!tpb]
    \centering
    \subfigure[]{\includegraphics[width=0.45\textwidth]{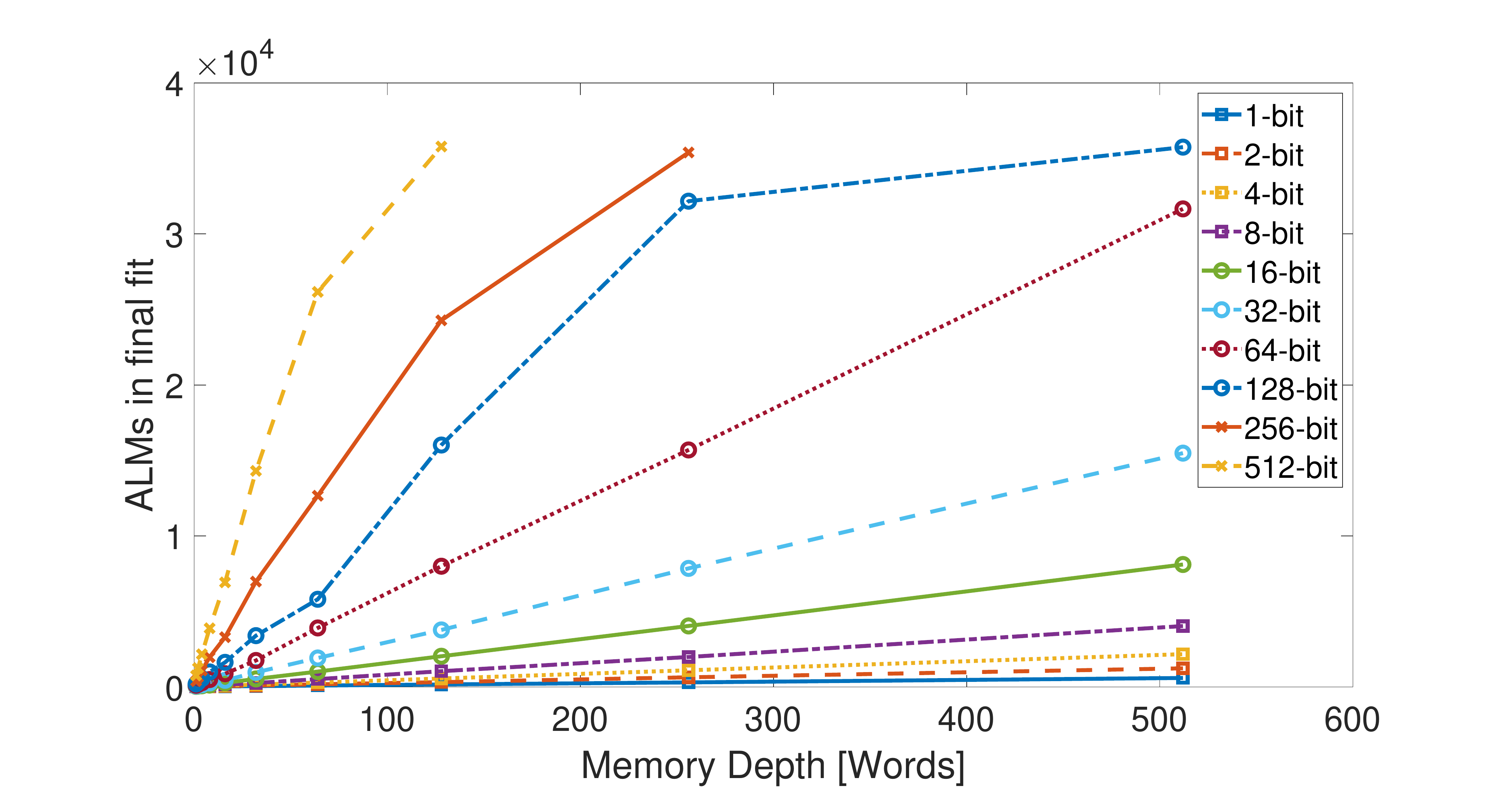}}
    \subfigure[]{\includegraphics[width=0.45\textwidth]{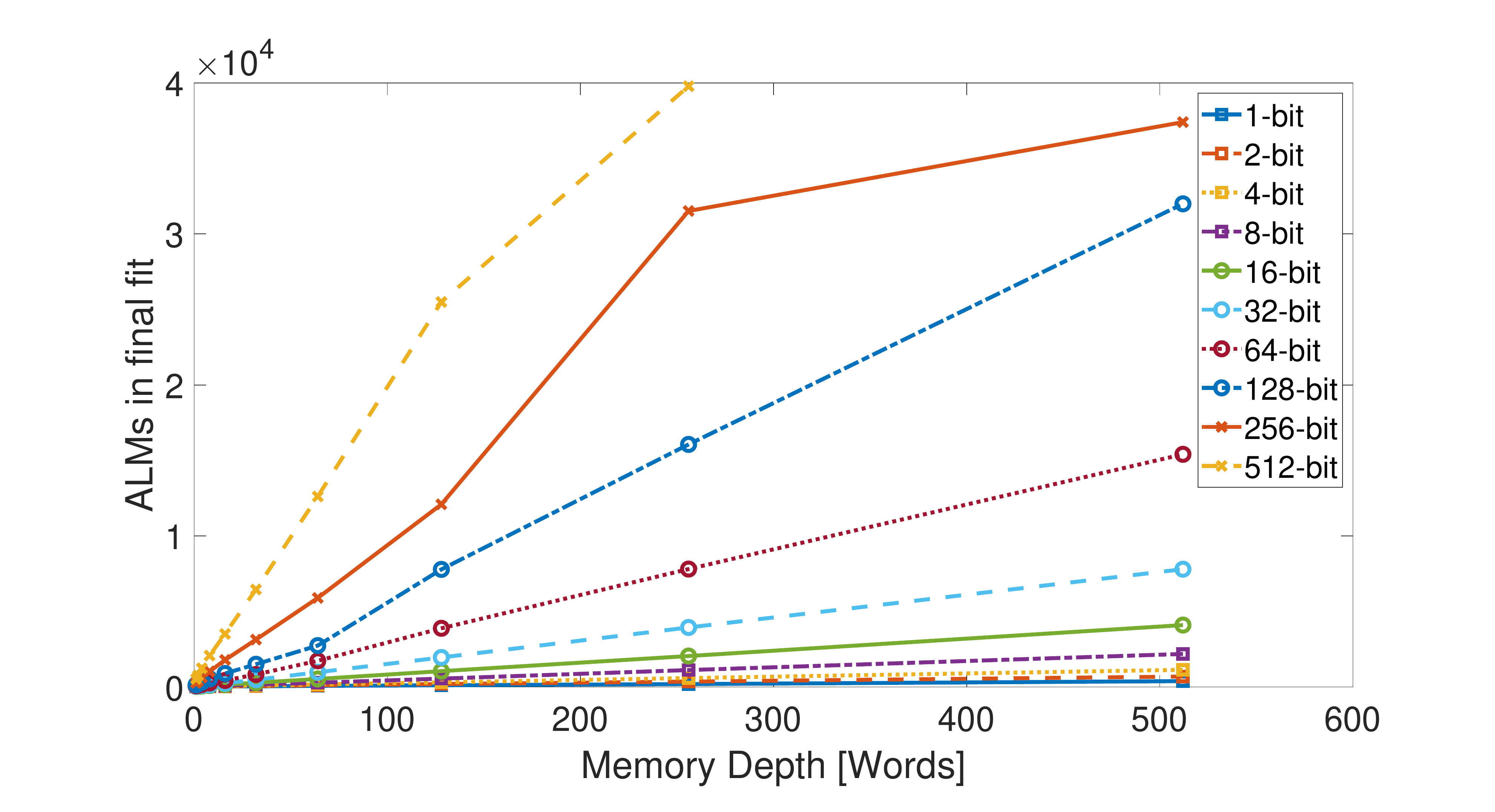}}
    \caption{ALMs used in final fit (Total ALMs less ALMs recovered from dense packing) for global memory module only. (a)~Global memory has a registered output. (b)~Global memory has no registered output.}
    \label{fig:global_no_target_alm}
\end{figure}

\begin{figure}[!tpb]
    \centering
    \subfigure[]{\includegraphics[width=0.45\textwidth]{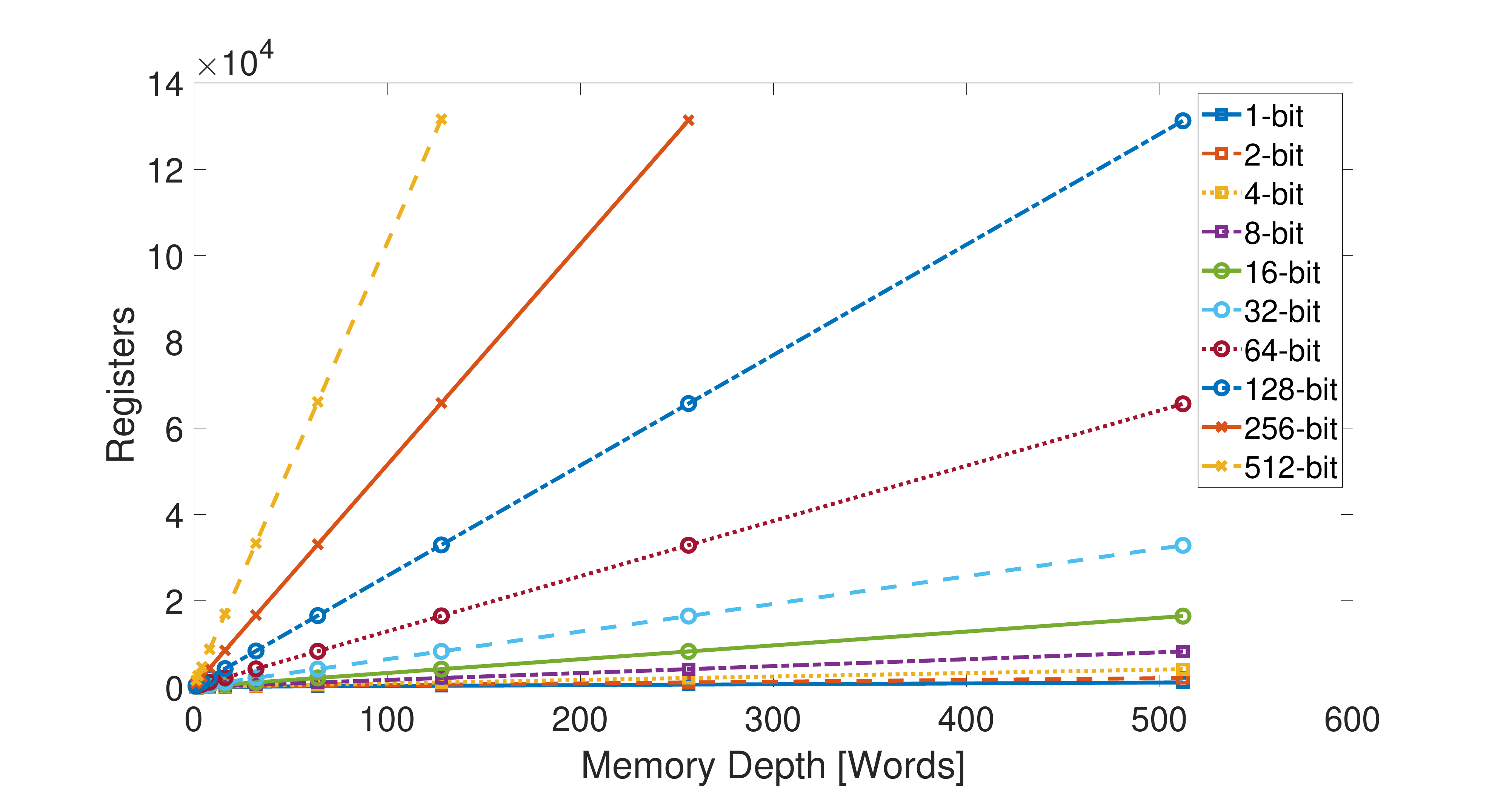}}
    \subfigure[]{\includegraphics[width=0.45\textwidth]{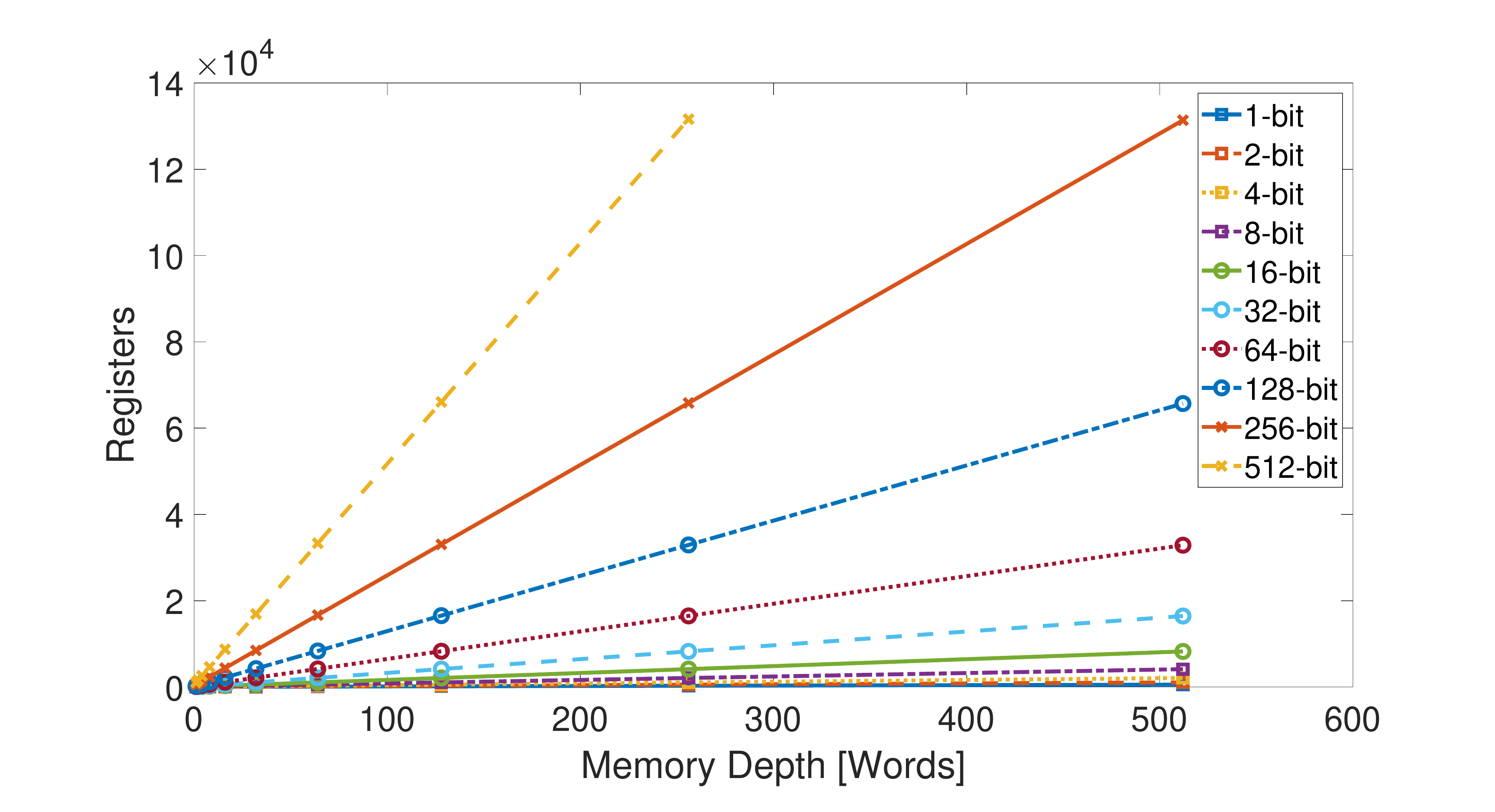}}
    \caption{Dedicated logic registers for global memory module only. (a)~Global memory has a registered output. (b)~Global memory has no registered output.}
    \label{fig:global_no_target_reg}
\end{figure}

\begin{figure}[!tpb]
    \centering
    \subfigure[]{\includegraphics[width=0.45\textwidth]{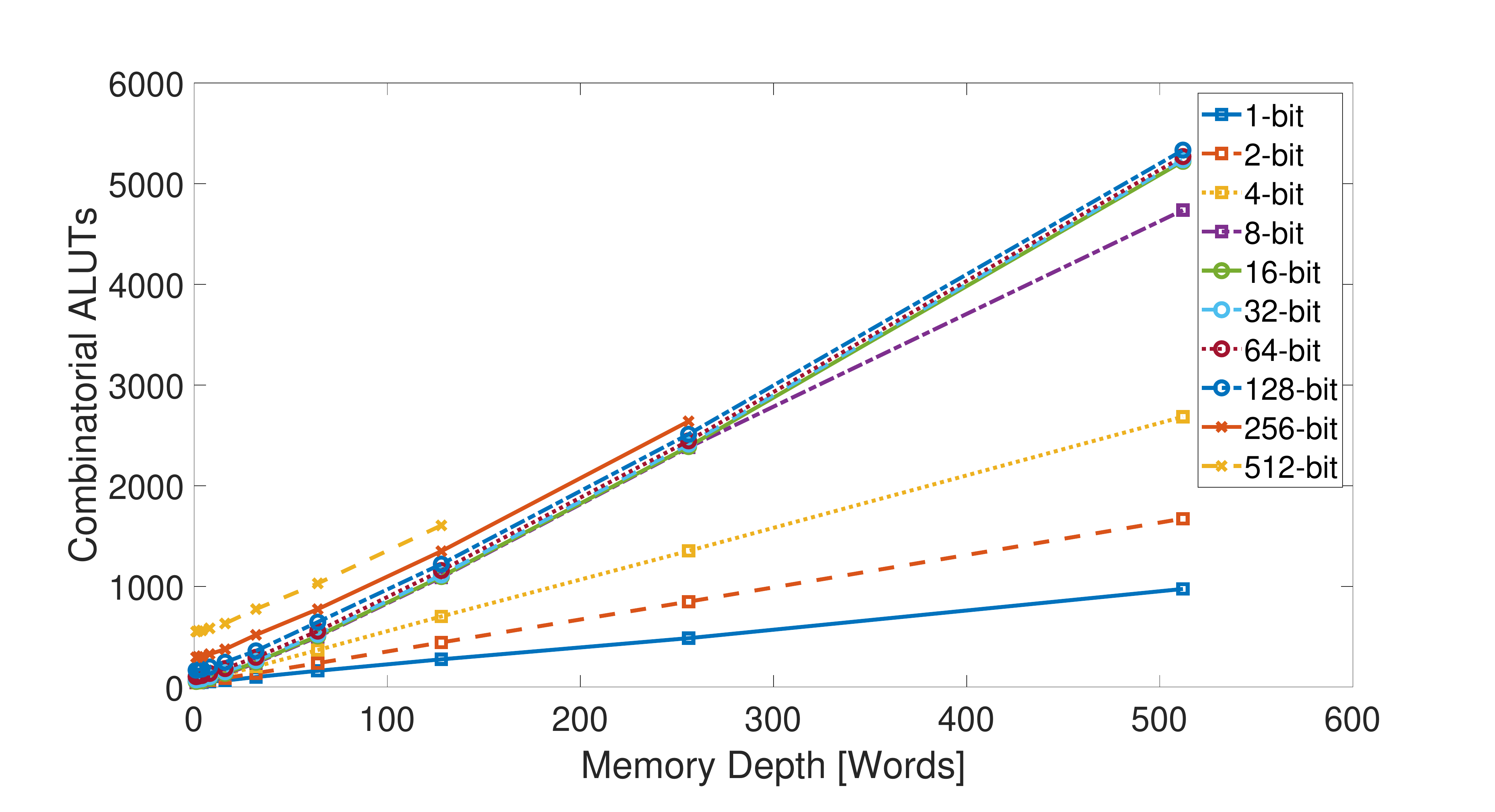}}
    \subfigure[]{\includegraphics[width=0.45\textwidth]{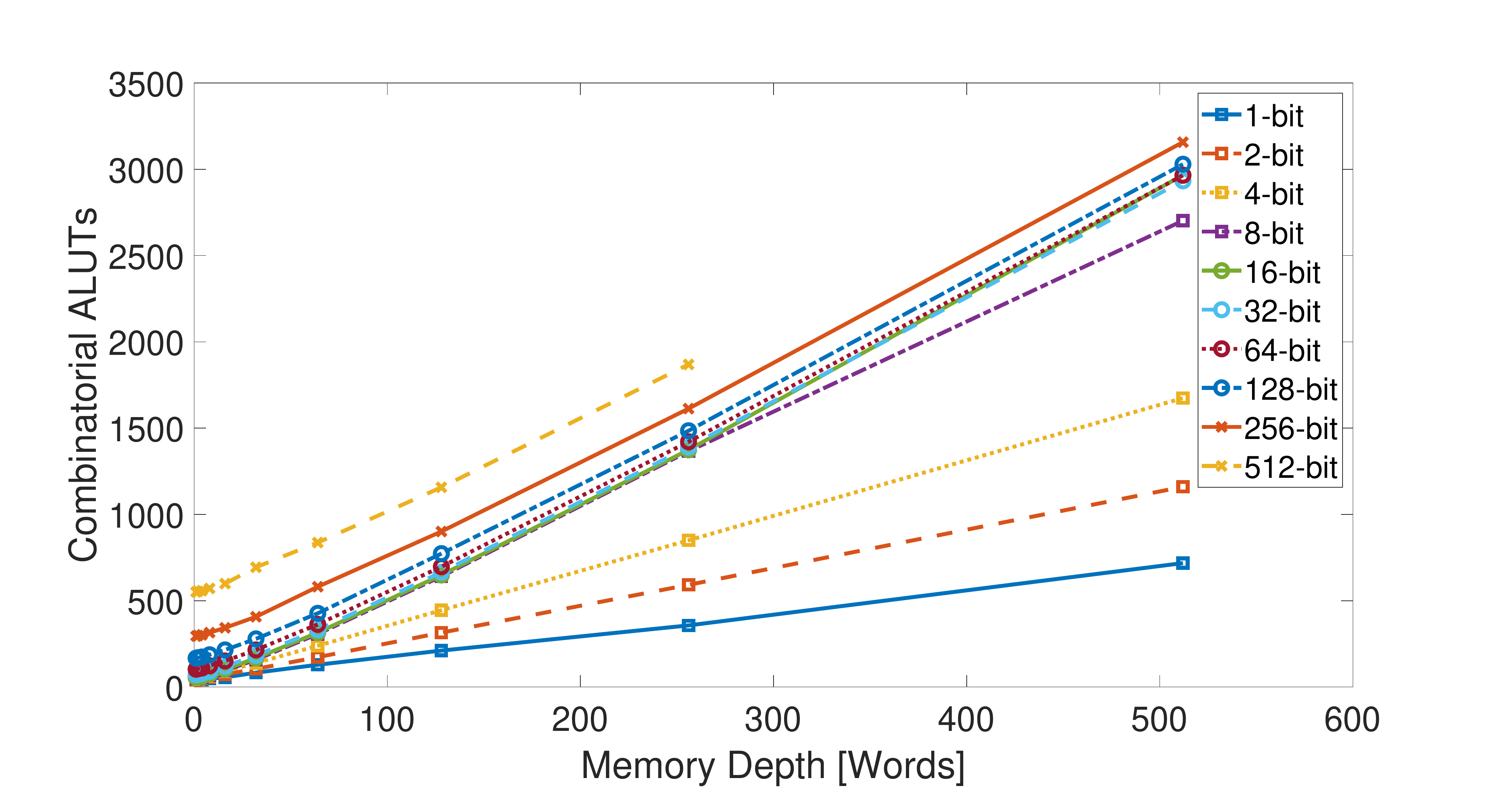}}
    \caption{Combinatorial ALUTs for global memory module only. (a)~Global memory has a registered output. (b)~Global memory has no registered output.}
    \label{fig:global_no_target_alut}
\end{figure}

\begin{figure}[!tpb]
    \centering
    \includegraphics[width=0.8\textwidth]{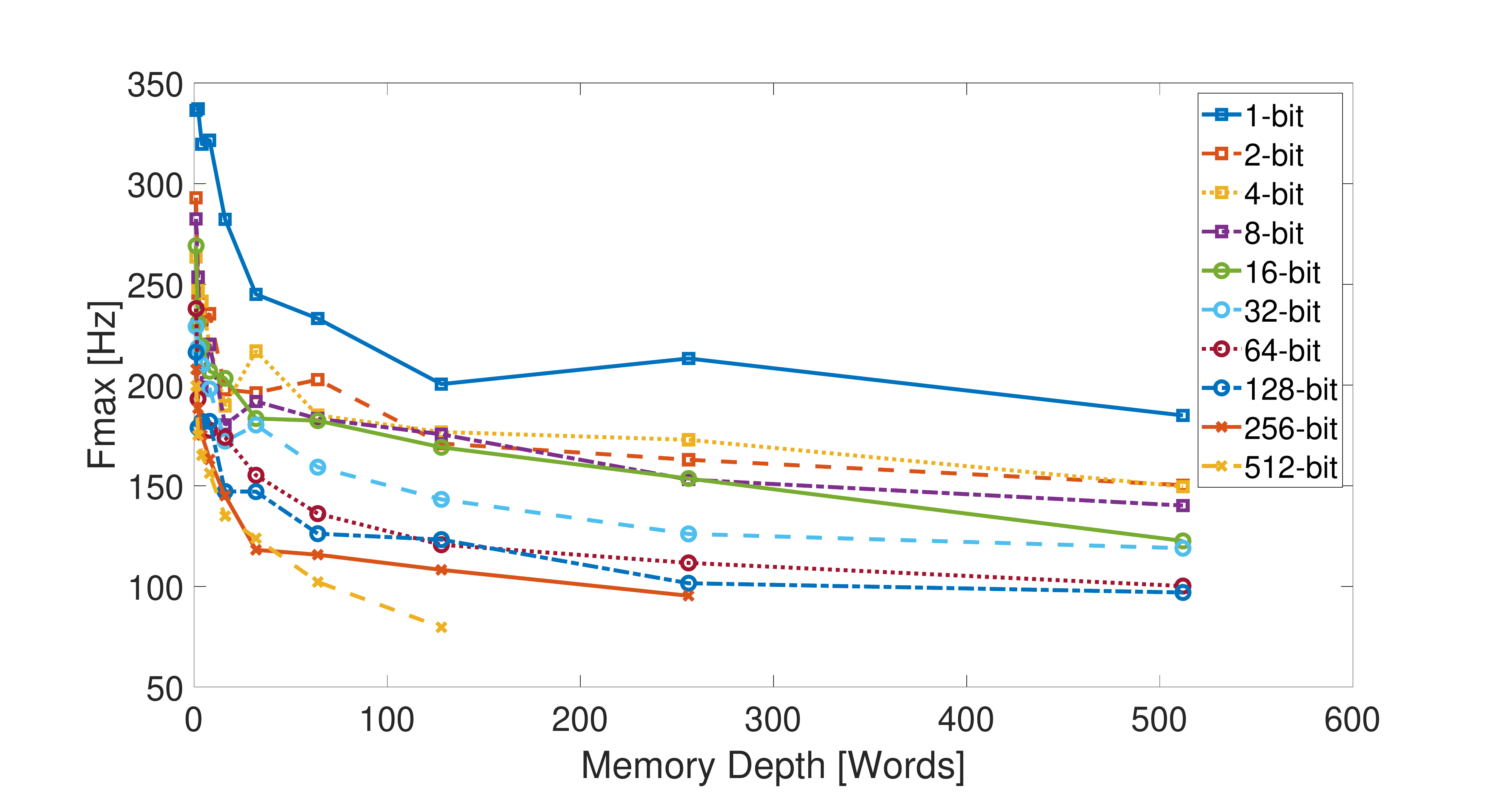}
    \caption{Maximum operating frequency (die temperature 85$^{\circ}$C) for global memory module only. Data only given for global memory with registered output.}
    \label{fig:global_no_target_fmax}
\end{figure}

Figures~\ref{fig:global_no_target_alm} to \ref{fig:global_no_target_fmax} show key metrics for an implementation of a global memory system. The global memory system contains the decoding logic for writing to the memory, the memory, and the output stage that would be connected to the rest of the design. These figures do not include the resource consumption of slave modules where the settings would be used and any clock domain crossing logic that may be implemented.\par

ALMs (Intel) --- similar to Configurable Logic Blocks (CLB) (Xilinx) --- contain a number of resources, typically (A)LUTs, adders, multiplexers, routing logic, and registers~
\cite{IntelALMWP}. From fig.~\ref{fig:global_no_target_alm} it is shown that adding a register stage to the output of the memory significantly increases the number of ALMs needed for implementation; for instance, in this case, 128 512-bit words with a final register stage require just over 10,000 (10,292.6) more ALMs for implementation --- approximately an extra 40\%. Similarly, the number of dedicated registers (fig.~\ref{fig:global_no_target_reg}) requires an extra 65,536 dedicated logic registers --- an approximately 100\% increase in resource. Again, the number of ALUTs, fig.~\ref{fig:global_no_target_alut}, has also increased by approximately 40\%. This is to be expected since the implementation shown in subfigures (a) of figs.~\ref{fig:global_no_target_alm} to \ref{fig:global_no_target_fmax} have an extra register stage per bit of the memory map at the output. 

This is an obvious draw back in terms of resource consumption. However, the accompanying benefits of the extra register stage is that the length of the routing between the memory and the target can now be broken down using the extra register stage. This manifests itself in an increase in operating frequency for the design. Figure~\ref{fig:global_no_target_fmax} shows the maximum operating frequency of the implementation that uses an extra register. While synthesising just the memory module itself we are unable to provide $f_{\text{max}}$ figures when there is no additional output register because there are no valid paths (paths between two flip-flops) for which the timing analyzer (TimeQuest) can operate.\par
 
\subsection{Global configuration with targets}
\label{sec:globaltargets}

In sect.~\ref{sec:global} the resource consumption for the memory decode logic and memory itself are shown. However, this is only half the story for a design that uses a global set of memory where entries are propagated out to other areas of the design. In this section we take a global memory system that a global memory of 256 32-bit words and propagates these out to a slave module with a varying number of configuration registers in the slave module. In addition, designs that use a combination of output registers on the global memory map, clock domain crossing registers (synchronisation chain length is 2 registers) and final location registers are examined. \par

\begin{figure}[!tpb]
    \centering
    \includegraphics[width=0.8\textwidth]{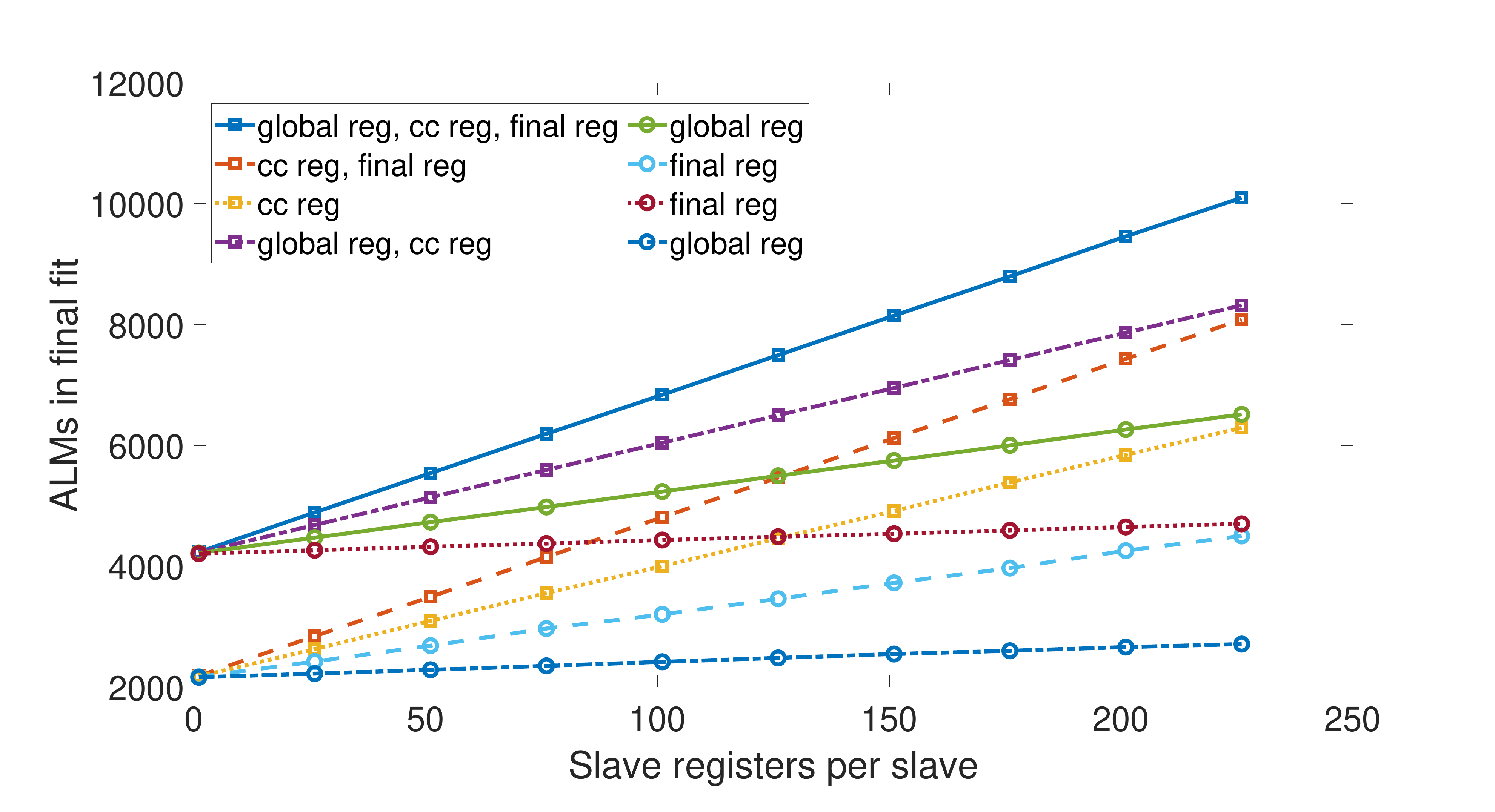}
    \caption{ALM consumption of global memory architecture with a single slave module using a variety of configuration registers and routing registers.}
    \label{fig:global_targets_alms}
\end{figure}

\begin{figure}[!tpb]
    \centering
    \includegraphics[width=0.8\textwidth]{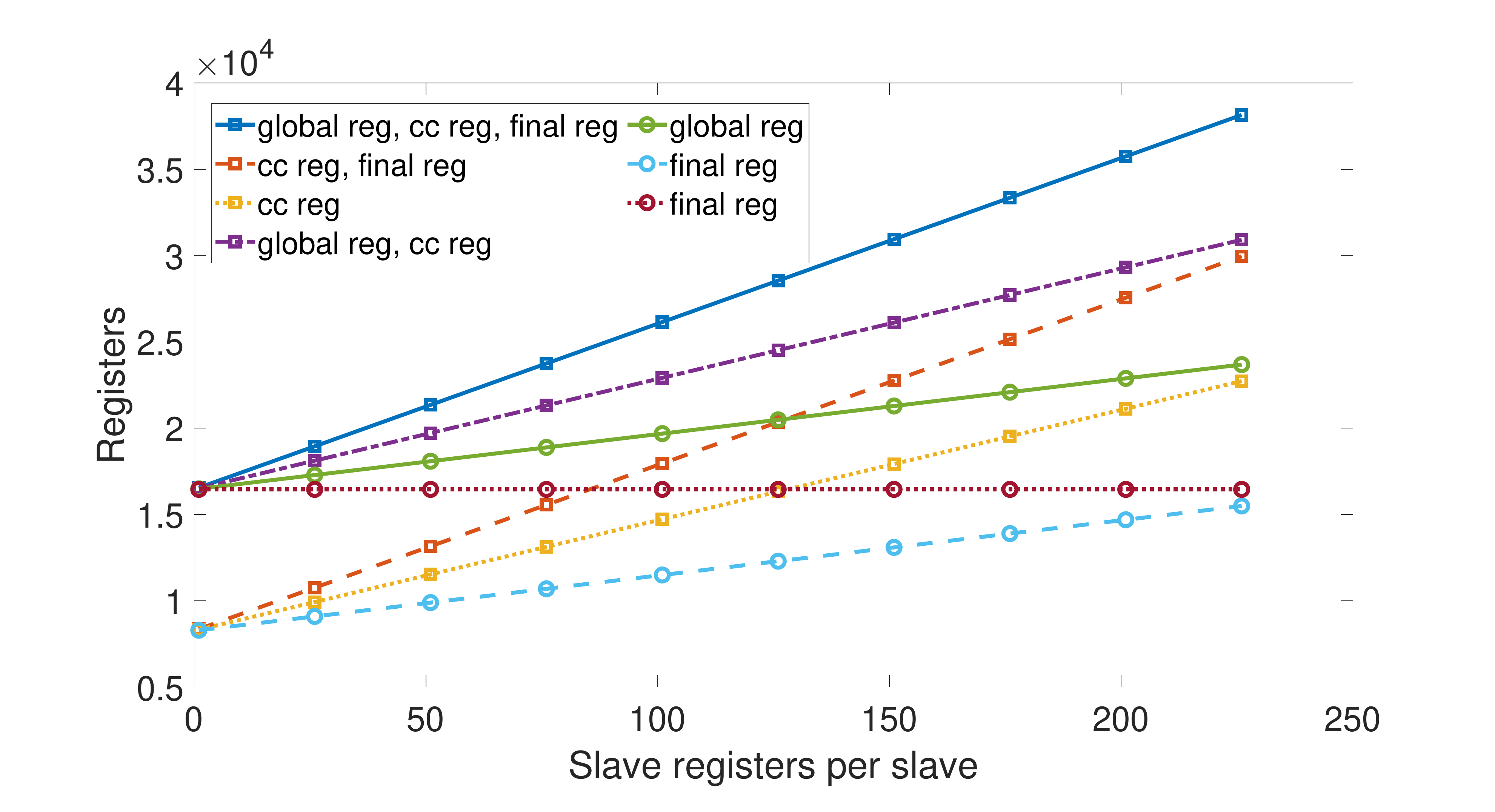}
    \caption{Register consumption of global memory architecture with a single slave module using a variety of configuration registers and routing registers.}
    \label{fig:global_targets_reg}
\end{figure}

\begin{figure}[!tpb]
    \centering
    \includegraphics[width=0.8\textwidth]{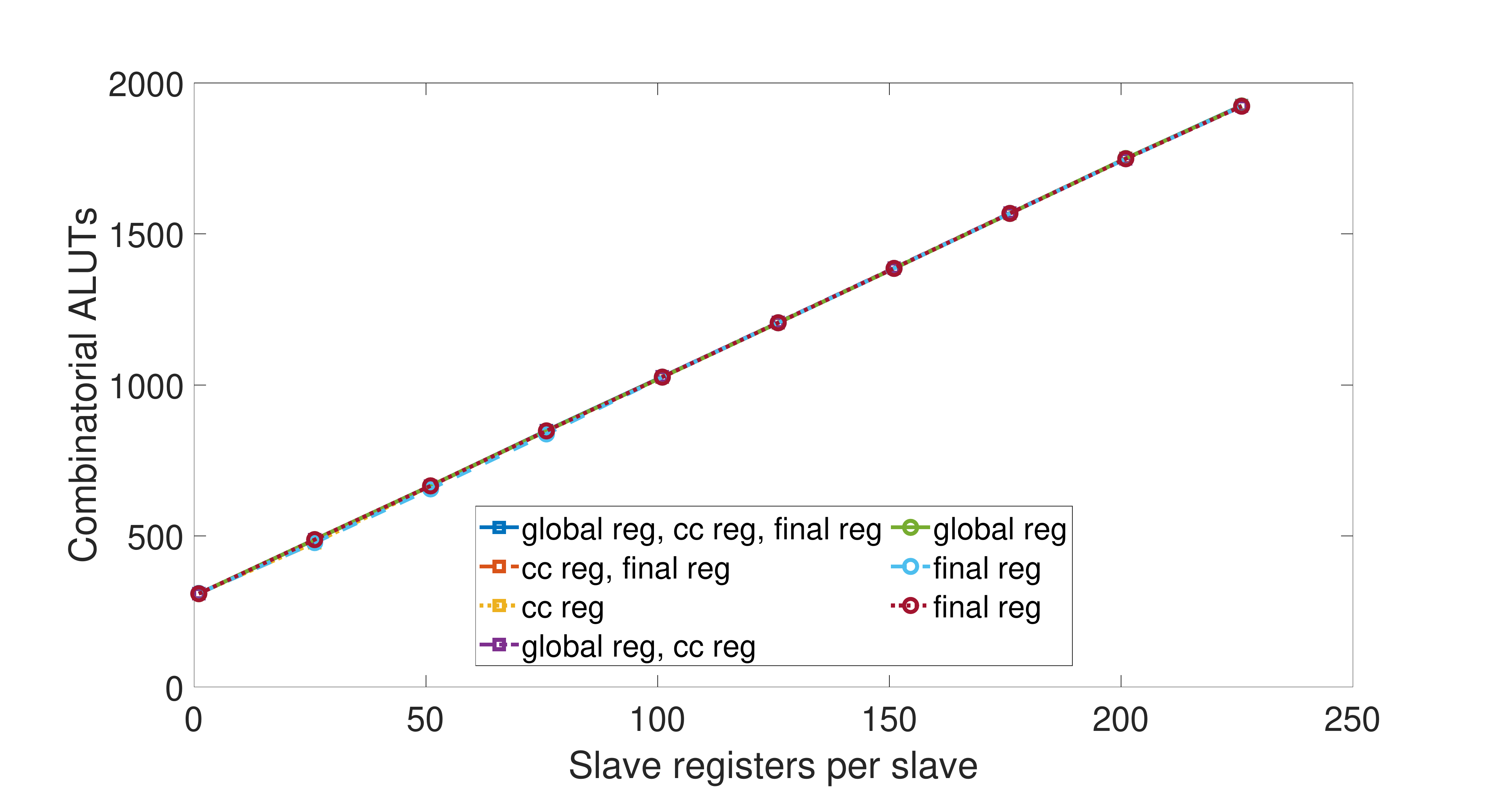}
    \caption{ALUT consumption of global memory architecture with a single slave module using a variety of configuration registers and routing registers.}
    \label{fig:global_targets_aluts}
\end{figure}

Figure~\ref{fig:global_targets_alms} is the after fitting ALM requirements, fig.~\ref{fig:global_targets_reg} is the after fitting register requirements, and fig.~\ref{fig:global_targets_aluts} is the after fitting ALUT requirements for each configuration of the global memory map architecture. As is expected, increasing the number of target registers linearly increases the requirement of each resource. Designs with a greater number of register stages (post global map register, clock domain crossing synchronisation chain registers and destination registers) significantly increases the resource requirements compared to design with fewer register stages. 10099.1\,ALMs, 38146\,registers, and 1925\,ALUTs for a design with 226 configuration registers and the maximum number of routing register stages compared to 2710.5\,ALMs, 8258\,registers, and 1913\,ALUTs for a design with the same number of configuration registers but no register stages to break down the length of the routing. The more crowded a design becomes, the greater the impact of removing the routing registers has on the maximum speed of a path. \par

\subsection{Distributed configuration}

\begin{figure}[!tpb]
    \centering
    \includegraphics[width=0.8\textwidth]{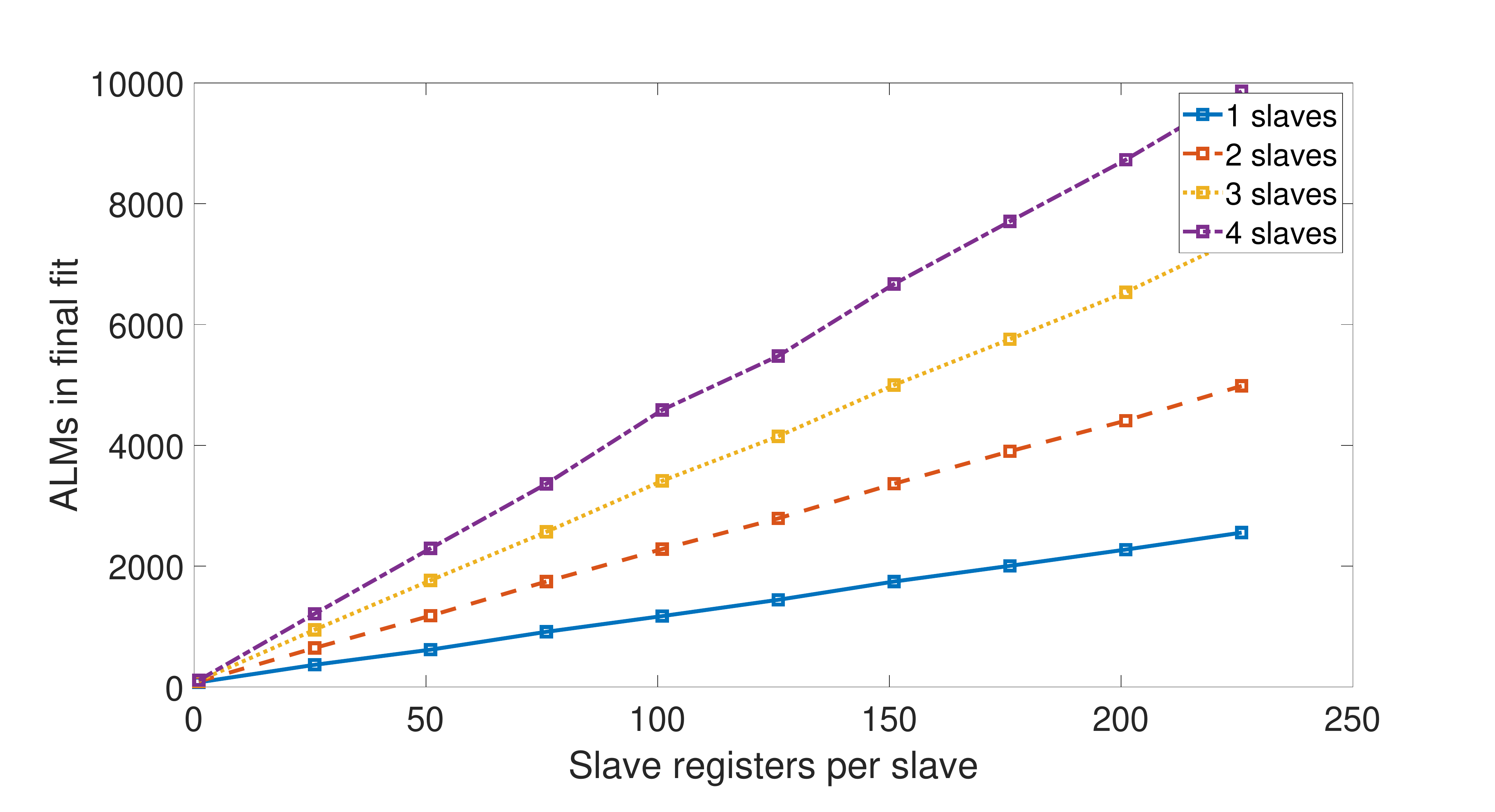}
    \caption{ALM consumption of distributed memory architecture with 1 to 4 slave module(s) and a variety of configuration registers.}
    \label{fig:distri_targets_alms}
\end{figure}

\begin{figure}[!tpb]
    \centering
    \includegraphics[width=0.8\textwidth]{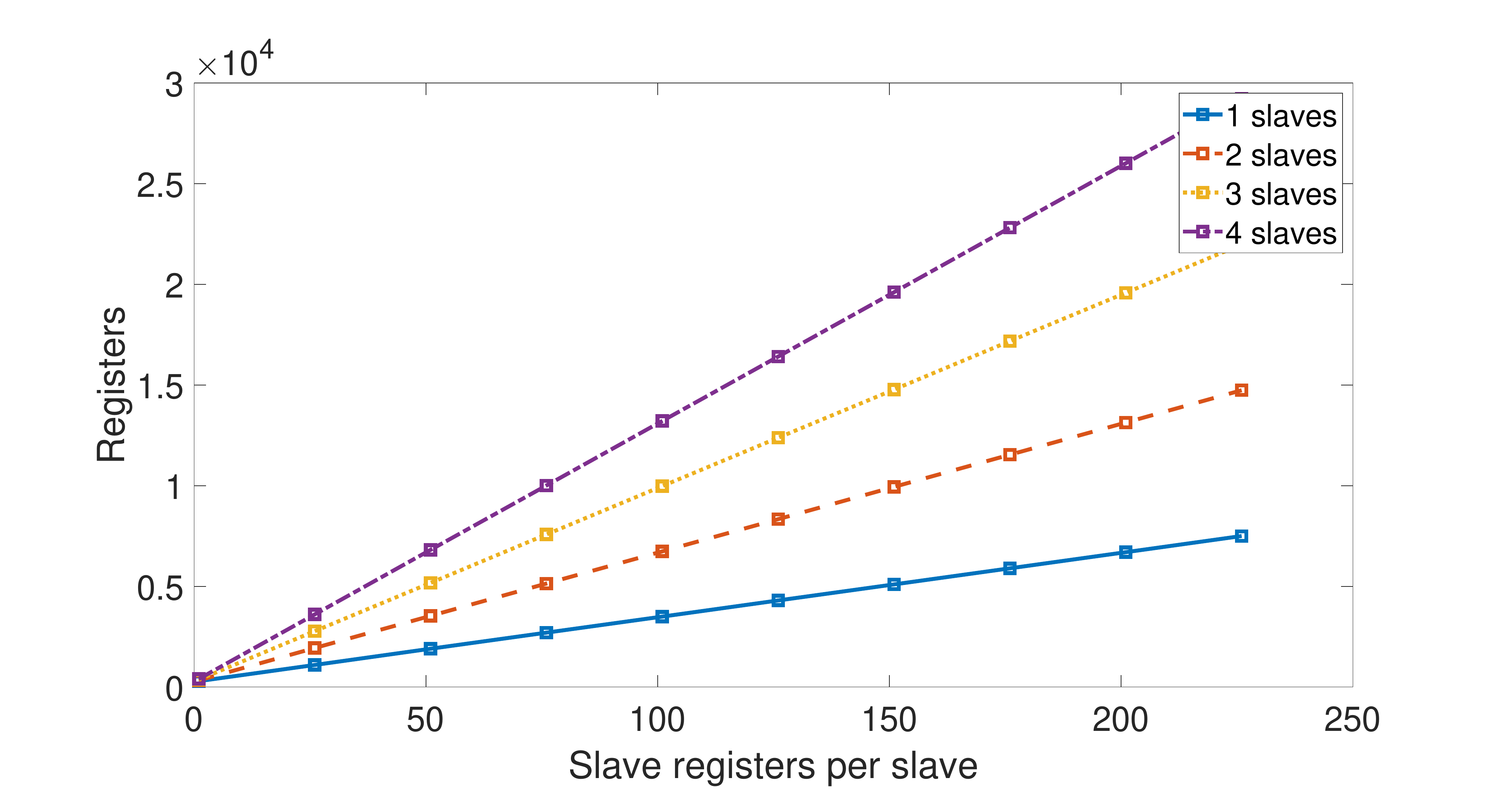}
    \caption{Register consumption of distributed memory architecture with 1 to 4 slave module(s) and a variety of configuration registers.}
    \label{fig:distri_targets_reg}
\end{figure}

\begin{figure}[!tpb]
    \centering
    \includegraphics[width=0.8\textwidth]{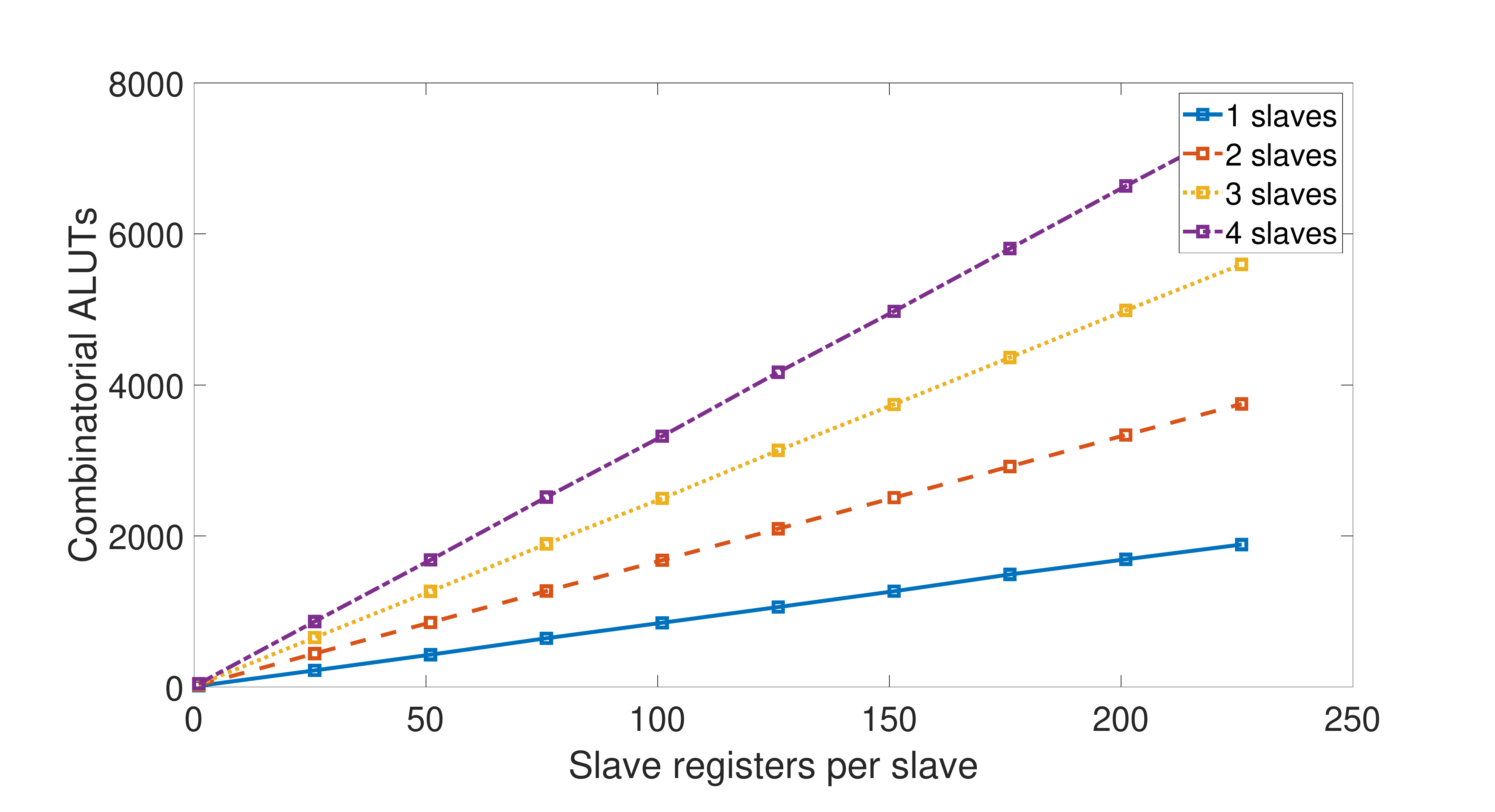}
    \caption{ALUT consumption of distributed memory architecture with 1 to 4 slave module(s) and a variety of configuration registers.}
    \label{fig:distri_targets_aluts}
\end{figure}

The resources required for the distributed configuration memory architecture, shown in figs.~\ref{fig:distri_targets_alms} to \ref{fig:distri_targets_aluts} are considerably lower than the global memory architecture. The graphs shown here are for implementations with a number of slave modules (1 to 4) each implementation varies the number of configurations per slave. For comparisons numbers from the `1 slave' implementations can be mapped to the results given in sect.~\ref{sec:globaltargets}. \par

The resources used for a distributed configuration memory implementation using 226 target registers per slave are: 2556.0\,ALMs, 7499\,registers, and 1887\,ALUTs. That is 25\% of the ALMs, 20\% of the registers used in the global design with maximum routing register. A significant cost saving. Increasing the number of slaves in the design has a linear effect on the resource cost. \par

\subsection{Operating frequency}

\begin{figure}[!tpb]
    \centering
    \subfigure[]{\includegraphics[width=0.45\textwidth]{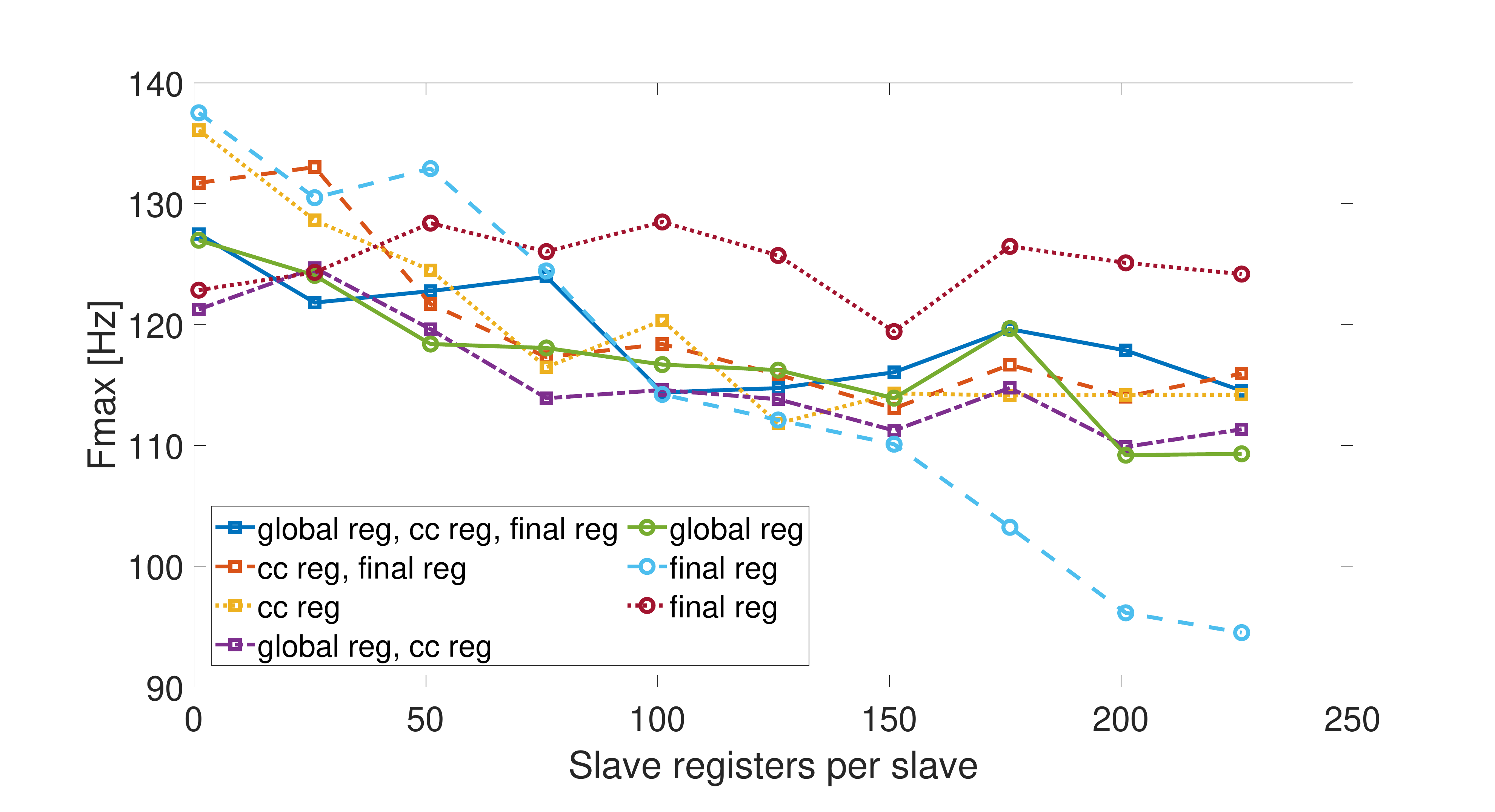}}
    \subfigure[]{\includegraphics[width=0.45\textwidth]{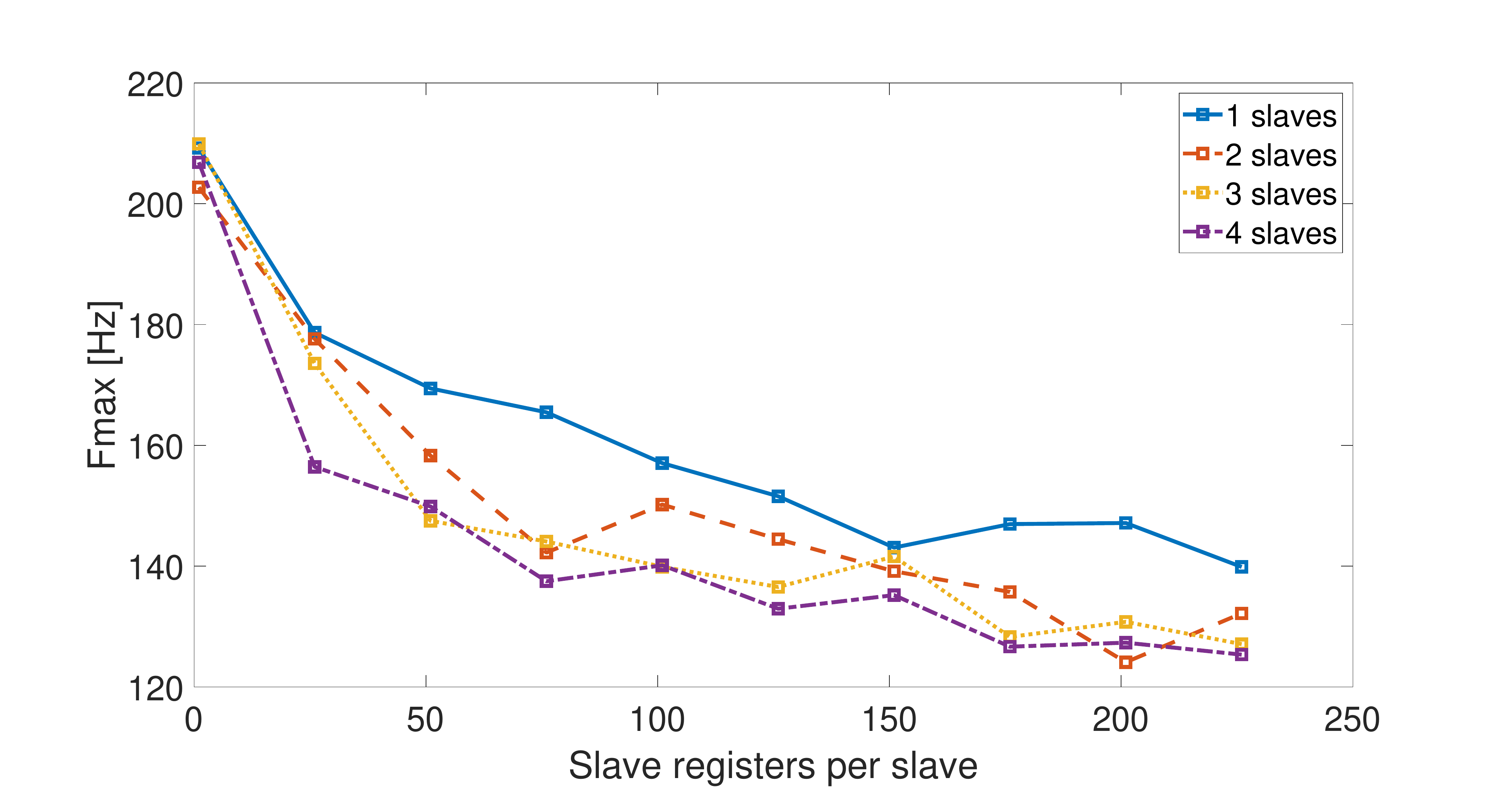}}
    \caption{$f_{\text{max}}$ for configuration memory architectures. (a)~Global memory architecture. (b)~distributed memory architecture.}
    \label{fig:targets_fmax}
\end{figure}

Figure~\ref{fig:targets_fmax} shows that the maximum operating frequency of a design is also influenced by the topology of the configuration architecture. A global memory architecture achieves a maximum $f_{\text{max}}$ of just shy of 140\,MHz compared to the approximate 210\,MHz of the distributed memory architecture.

\section{Conclusions}
\label{sec:conclusions}

In this paper it has been shown that there are a number of ways to achieve the implementation of configuration registers in an FPGA design. In this paper we proposed a global memory architecture and a distributed memory architecture, for completeness the global memory architecture was presented with combinations of register stages and clock domain crossing registers. It has been shown that the distributed architecture has a much lower resource cost for ALMs and registers (as small as 25\% and 20\% respectively for a design using 226 32-bit configuration registers). It has further been shown that there is a disparity in the maximum operating frequency between the designs with the distributed memory architecture achieving a higher maximum operating frequency. \par 
Aside from the reduction in resource cost between the different architectures, the distributed memory architecture uses a common configuration bus that is independent of the number of target registers and their width. The uniformity of the configuration bus opens up the ability to implement the configuration system in a partially re-configurable FPGA design, where the configuration bus can be connected to any re-configurable design without penalty. Similarly, the architecture of the configuration bus is not liable to mis-sampling when crossing clock domains. It is set only when the slave module reports it is safe to change the settings.  

\bibliographystyle{plain}
\bibliography{references, references_dpr, references_fpga,references_distributed_memories}

\end{document}